\documentstyle[aps,prd]{revtex}
\newcommand{\be}{\begin{equation}}
\newcommand{\ee}{\end{equation}}
\newcommand{\ba}{\begin{eqnarray}}
\newcommand{\ea}{\end{eqnarray}}

\newcommand{\ib}{Ib\'a\~{n}ez {\it et al.}\,}
\newcommand{\no}{\noindent}

\begin{document}

\title{Reply to ``On scaling solutions with a dissipative fluid"}

\author{
Luis P. Chimento\footnote{Electronic address:
chimento@df.uba.ar}  and  Alejandro S. Jakubi\footnote{
Electronic address: jakubi@df.uba.ar}}
\address{Departamento de F\'{\i}sica, Universidad de
Buenos Aires, 1428~Buenos Aires, Argentina}
\author{Diego Pav\'on\footnote{Electronic address: 
diego.pavon@uab.es}}
\address{
Departamento de F\'{\i}sica, Universidad Aut\'onoma
de Barcelona, 08193 Bellaterra, Spain}

\date{\today}

\maketitle

\begin{abstract}
In this paper we show that the claims in \cite{ibanez} related to
our analysis in \cite{enlarged} are wrong.
\pacs{98.80.Hw}
\end{abstract}

As is well known, current observational evidence strongly favors an 
accelerating and spatially flat Friedmann--Robertson--Walker universe 
\cite{schmidt}, \cite{debernardis}. Since normal matter obeys the strong
energy condition and cannot drive accelerated expansion, recourse is often
made either to a small cosmological constant or an almost evenly
distributed source of energy -termed ``quintessence"- in the form of a
self--interacting scalar field with equation of state $p_{\phi} =
(\gamma_{\phi} - 1) \rho_{\phi}$ (where $ 0\leq \gamma_{\phi} <1$) such
that it provides a negative pressure high enough to render the
deceleration parameter negative (i.e., $q \equiv - \ddot{a}/(a H^{2}) <
0$), see e.g. \cite{caldwell}. As usual, $a$ denotes the scale factor of
the FRW metric and $H \equiv \dot{a}/a$ the Hubble factor. Owing to the
fact that in general the scalar field and matter energy are expected to
scale differently with expansion, the question arises: ``why the ratio
between matter and quintessence energies should be of the same order
precisely today?" Put another way, ``where the relationship
$(\Omega_{m}/\Omega_{\phi})_{0} \simeq {\cal O}(1)$ comes from?" This is
the so-called {\it coincidence problem} \cite{steinhardt}. Here the zero
subindex means present time and the dimensionless density parameters of
matter and quintessence are defined by \\ \[ \Omega_{m} \equiv
\frac{\rho_{m}}{3H^{2}}, \qquad \Omega_{\phi} \equiv
\frac{\rho_{\phi}}{3H^{2}} \qquad (c= 8\pi G = 1), \] \\ respectively.
Note that the cosmological constant case is recovered by setting
$\gamma_{\phi} = 0$.

Obviously, solving the coincidence problem in an accelerating universe
amounts to show that the ratio $\Omega_{m}/\Omega_{\phi}$ tends to a
constant for large times with $q <0$. Then, one may choose the free
parameters of the model to fix the constant ratio to order unity. In a
recent paper we demonstrated that the coincidence problem and an
accelerated expansion phase of FRW cosmologies cannot be  {\em
simultaneously} addressed simply by the combined effect of a perfect
matter fluid and a quintessence scalar field. Nonetheless, if the matter
fluid is not perfect but dissipative, both problems can find a
simultaneous solution for FRW spatially flat and open FRW universes
\cite{enlarged}.

A key point in our derivation was the stationary condition

\be
\gamma_{m} + \frac{\pi}{\rho_{m}} = \gamma_{\phi}
= - \frac{2 \dot{H}}{3H^{2}}, 
\label{attractor} 
\ee

\noindent
as well as the violation of the strong energy condition

\be
\pi < \left(\textstyle{2\over{3}} - \gamma_{m} \right) \rho_{m}
\label{vSEC}
\ee

\noindent (cf. \cite{enlarged}), where, as usual, $\gamma_{m}$ denotes the
baryotropic index of matter, i.e., $p_{m} = (\gamma_{m} -1) \rho_{m}$ thereby
$1 \leq \gamma_{m} \leq 2$. Likewise $\pi$ stands for the dissipative stress
which is negative for expanding universes. It may be associated with particle
production \cite{zeldovich}, may be understood as frictional effects arising
in mixtures \cite{udey} or even  model other kinds of sources (e.g. a string
dominated universe as described by Turok \cite{turok}, or a scalar field).

Equation (\ref{attractor}) expresses the condition for 
$\Omega_{m}$ and $\Omega_{\phi}$ to be constants. It can be 
straightforwardly derived by combining the conservation 
equations for the matter fluid and quintessence field 
\\
\be
\dot{\rho}_{m}+ 3 \left(\gamma_{m} + \frac{\pi}{\rho_{m}}\right)
\rho_{m} H = 0, \quad \mbox{and} \qquad
\dot{\rho}_{\phi}+ 3 \gamma_{\phi} \rho_{\phi} H = 0,
\label{conserv}
\ee

\noindent with the definitions of the density parameters $\Omega_{m}$ and
$\Omega_{\phi}$, respectively. Thus, the condition $\dot{\Omega}_{m} =
\dot{\Omega}_{\phi} = 0$ translates into (\ref{attractor}). Equation
(\ref{vSEC}) expresses the asymptotic stability of the solution
$\Omega=\Omega_m+\Omega_\phi=1$. By slightly perturbing the solutions
$\Omega_{i} = $ constant ($i = m, \phi$) one finds they are stable for
spatially flat and open accelerating universes -see Eqs (18) and (22) of
\cite{enlarged}. At this stage we would like to emphasize two points, namely:
(i) our derivation shows that a dissipative stress negative enough to satisfy
the Eqs. (\ref{attractor}) and (\ref{vSEC}) is required to provide a
coincidence solving attractor solution. This derivation does not hinge on the
specific choice of the potential of quintessence field. (ii) Our proof is
meant just for universes under accelerated expansion, i.e., for $q =
\textstyle{3\over{2}}\gamma_{\phi} - 1 < 0 $. Clearly, the conditions
expressed by Eqs. (\ref{attractor}) and (\ref{vSEC}) define the set of
transport equations for $\pi$ or, equivalently, the set of expressions of the
bulk viscosity coefficient, leading to accelerated stable solutions (see \S 
IV of \cite{enlarged} for some examples).

However, recently \ib  studied the autonomous system of equations of a FRW
universe filled with a dissipative matter fluid and a self--interacting
scalar field $V(\phi) \propto \mbox{exp} (k \phi)$ and found stable and
unstable equilibrium points with $\Omega_{i} = $ constant \cite{ibanez}.
The stability of these points seems to depend on the specific equation of
state of $\pi$ as well as on the values assumed by different parameters of
their model. Since this outcome looks at variance with point (i) of the
precedent paragraph \ib erroneously claimed to have found examples that
contradict our findings. The fact is, however, they overlooked point (ii):
all the unstable solutions in \cite{ibanez} correspond to
non--accelerating universes.  To be specific: the equilibrium point of 
Eqs. (29) has $q > 0$ (cfr. Eq. (32)). The set of equilibrium points 
associated to a massless scalar field ($\Gamma =0$) have also $q > 0$. 
Likewise, the equilibrium point of Eqs. (40)-(42) is either stable and 
accelerating or unstable and decelerating (see Eq. (48) and Fig. 4). Again, 
the equilibrium point of Eqs. (49)-(52) is non-accelerating (see Eq. (53)).
Finally, the equilibrium point (54)-(56) has $q >0$.

Therefore they do not invalidate in any way whatsoever the findings of 
\cite{enlarged}. We believe this is more than enough to dismiss the 
claims of \ib However, we have found inconsistencies in their analysis. 
In the remaining of this paper we concentrate in bringing them 
to the fore.

The starting equations of \ib are: 
\begin{eqnarray} 
\dot H        &= &
-H^{2}-\frac{1}{6}\left(\rho_{m}+3p_{m}+3\pi+2\dot\phi^{2}
                   -2V(\phi)\right)\\ 
3H^2          &= & \rho_{m}+\frac{1}{2}\dot\phi^{2}+V(\phi)-\frac{3K}{a^2}
\qquad (K=\pm 1,0)\\ \dot\rho_{m}  &= & -3H(\rho_{m}+p_{m}+\pi)\\
\ddot\phi &= & -3H\dot\phi-\frac{dV(\phi)}{d\phi}. \label{KG}
\end{eqnarray}

\no The dissipative pressure $\pi$ is assumed to satisfy the truncated
Israel-Stewart equation \cite{werner}
\\
\be
\label{trunc}
\pi+\tau\dot\pi =-3 \zeta H,
\ee
\\
where $\zeta$ is the coefficient of bulk viscosity and $\tau$  the
relaxation time ($\zeta>0, \tau>0$). Strictly speaking, Eq. (\ref{trunc})
is valid only when the fluid is close to equilibrium. However, let us
assume it holds even when the fluid is far from equilibrium. (A more
rigorous and comprehensive analysis would make use of the full transport
equation of the Israel-Stewart theory.) \ib also assumed the linear
baryotropic equation of state $p_{m} =(\gamma_{m}-1)\rho_{m}$, but with
$\gamma_{m} = $ constant, and two  different relations for the coefficient
of bulk viscosity and relaxation times. To translate (3)-(\ref{trunc})
into an autonomous system  of differential equations \ib  introduced 
the set of variables 
\\
\begin{eqnarray} \Omega_{m} & = & \frac{\rho_{m}}{3H^2},\qquad
\Sigma=\frac{\pi}{H^2}, \qquad \Psi=
\frac{1}{\sqrt{6}}\frac{\dot\phi}{H},\nonumber\\ \Gamma & = &
\frac{1}{3}\frac{V(\phi)}{H^2},\qquad h=H^{1-2n}, \qquad n\ne
\frac{1}{2},\label{vars} \end{eqnarray} \\ as well as a new time parameter
$\tau$ defined by \be \label{tau} d\tau = H(t)\;dt. \ee \\ The Friedmann
equation (4) now reads \be 1-\Omega_{m}^2-\Psi^2-\Gamma=-\frac{K}{H^2
a^2}, \label{F} \ee \\ and the autonomous system takes the form \\
\begin{eqnarray}\label{dO} \Omega'_{m} & = & \Omega_{m}
(-3\gamma_{m}+2x)-\Sigma\\ \label{sigma} \Sigma'     & = &
-9\Omega_{m}+\Sigma\left[-\frac{1}{\alpha}(3\Omega_{m})^{(1-n)}h+2x\right]
\\ \Psi'       & = & \Psi(x-3)-\frac{3k}{\sqrt{6}}\Gamma\\ \Gamma' & = &
\Gamma (2x+k\sqrt{6}\Psi)\\ 
           h' & = & -(1-2n)hx
\label{h} 
\end{eqnarray}
\\ 
where the prime denotes derivative with respect to $\tau$, and
$x \equiv 1+\textstyle{1\over{2}}(3\gamma_{m}-2)\Omega_{m}+
\textstyle{1\over{2}} \Sigma+2\Psi^2-\Gamma $. 

Note that Eq. (\ref{sigma}) is the truncated Israel-Stewart transport
equation, while Eq. (\ref{h}) is just the derivative of the change of
variables introduced in (\ref{vars}). In Eq. (\ref{sigma}), first case of \ib,
use has been made of the relationships $\zeta=\alpha \rho_m^n$, $\tau= \alpha
\rho_m^{n-1}$, introduced by Belinskii {\it et al.} \cite{belinskii}. In their
second case a somewhat modified relationships were used for $\zeta$ and $\tau$
(introduced by their Eqs.(37)), amounting to replace (\ref{sigma}) by a
similar equation.

The equilibrium points are found by setting $\Omega'_{m} = \Sigma'= \Psi'
= \Gamma' = h'= 0$. As a consequence, $\rho_{m}$, $\pi$, $\dot{\phi}$,
$V(\phi)$ and $H$ are constants there (use of the variables introduced in Eqs.
(\ref{vars}) and (\ref{tau}) excludes $H=0$). Then, in virtue of Eq. (\ref{KG})
we have $d V(\phi)/d \phi = 0$ and $\dot{\phi} = 0$, i.e., $V(\phi)$ must
have an extremum at the equilibrium points, and $\gamma_{\phi} = 0$. 
Potentials having this feature yield de Sitter solutions on the equilibrium
points and for them, as noted above,  we could replace the quintessential 
field with an effective cosmological constant.

~From Eq. (4), with $H$, $\rho$, $V(\phi)$ constants and $\dot{\phi} = 0$, we
get $ K = 0$. Moreover, from (5) it follows that $\pi = -\gamma_{m} \rho_{m}$.
This is independent of the specific form of $V(\phi)$. Notice that
this is a particular case of the result already found in \cite{enlarged}, as
follows from (\ref{attractor}) and the assumption $H = $ constant.

The setting $h' = 0$ by \ib (their Eq.(18)) leads to wrong conclusions. In 
the first case (Belinskii {\it et al.} relationships), they erroneously 
state that its solution is $h = 0$ rather than 
$h = -3\alpha(3\Omega_m)^n/\Sigma$  (with $x=0$), as follows from their 
Eqs.(15) and (19); and $h =$ constant in the second case. Besides, their 
relationship $h' = 0$ neither belongs to the set of Einstein equations 
nor describes any property of the sources of the gravitational field. 

Altogether, the work of \ib misinterpret our results (to say the least)
rendering the claims in \cite{ibanez} void. Further, even if the analysis 
of \ib were correct (that is not), it would fail to disqualify our findings in 
\cite{enlarged} as all their counterexamples correspond to 
non--accelerated cosmic expansions.

\section*{Acknowledgments}
LPC and ASJ thank the University of Buenos Aires for partial support under
project X-223. This work was partially supported by the Spanish Ministry of
Science and Technology under grant BFM 2000--0351-C03--01 and 2000--1322.

\end{document}